\documentstyle[12pt,a4]{article}
\begin{document}
\newcommand{\bm}{\bibitem}
\newcommand{\ud}{\bf}
\normalbaselineskip = 24 true pt
\normalbaselines
\bibliographystyle{unsrt}
\def\be {\begin{equation}}
\def\ee {\end{equation}}
\def\bea {\begin{eqnarray}}
\def\eea {\end{eqnarray}}
\def\et {{\it et al}}
\def\gf {{\gamma_5}}
\def\gmu {{\gamma_\mu}}
\def\tauiso {{\mbox{\boldmath $\tau$}}}
\thispagestyle{empty}
\addtolength{\topmargin}{-2cm}
\textheight 23.0  cm
\renewcommand{\thefootnote}{\fnsymbol{footnote}}
\begin{center}
{\LARGE Pion production in proton-proton collisions in a covariant one boson
exchange model\footnote{Work supported by GSI, Darmstadt,
KFA J\"ulich and JSPS.}}\\[1.0cm]
{\bf A. Engel$^a$,
A. K. Dutt-Mazumder$^b$, R. Shyam$^{b,c}$ and U. Mosel$^c$
}\\[2mm]
$^a${\it Department of Physics, Kyoto University, Kyoto 606-01, Japan}\\
$^b${\it Saha Institute of Nuclear Physics Calcutta - 700 064, India}\\
$^c${\it Institut f\"ur Theoretische Physik,Universit\"at Giessen,
D-35392 Giessen, Germany}\\[2 cm]
\end{center}
\renewcommand{\thefootnote}{\arabic{footnote}}
\begin{abstract}
Motivated by the
renewed interest in studying the pion production
on nuclei with
protons at few GeV incident energies,
we investigate the pion production in 
proton-proton collisions over an energy range of 300 $MeV$
to 2 $GeV$.
Starting from a realistic one-boson
exchange model with parameters
fitted to the amplitudes of the elastic nucleon-nucleon 
scattering, we perform fully covariant calculations for the total,
double and triple differential cross-sections of
the $p(p,n\pi^+)p$ and
$p(p,p\pi^0)p$ reactions. 
The calculations incorporate the exchange of $\pi,
\rho,\omega$ and $\sigma$ mesons and treat nucleon and delta isobar as
intermediate states. We obtain a reasonably good agreement with the
experimental data in the entire range of beam energies.
The form of the covariant delta propagator, the cut-off parameter
for the $\pi NN$ and $\pi N\Delta$ vertex form factors and the energy 
dependence of the delta isobar decay width is investigated.
\end{abstract}
\newpage
\section{Introduction}
The study of pion production in two nucleon collisions dates back to the mid
fifties when one of the earliest measurements of the $p(p,n\pi ^{+})p$
reaction was performed \cite {1}. This was followed by several
measurements of this reaction in 1960's and early seventies \cite{2}. These
experiments were usually conducted in emulsions or bubble chambers,
consequently the data invariably had poor statistics. However,
with the advent of accelerators
capable of producing intense high quality beams of protons
with energies up to a few GeV and
sophisticated detecting systems, it became possible to obtain good
quality data on the charged as well as neutral pion production in
proton-nucleon collisions at several beam energies \cite {3,4}.
Even the complete kinematical measurements of the three particles
in the final state can now be performed \cite {5,6,7,8}.
Furthermore, with improved polarisation techniques for  beams and targets,
the spin observables (asymmetries) are also being measured with
great accuracy \cite{6,7}.
These measurements provide a very stringent test
of the theoretical models of the
pion production which is the dominant inelastic channel in nucleon-nucleon
(NN) collisions.

A thorough investigation of the pion production in the elementary 
NN collisions within a fully relativistic model is
essential in the context of recent efforts to develop a covariant two
nucleon model (TNM) to describe the $A(p,\pi)B$ reactions \cite{9,10},
where several parameters have to be fixed from the study of the
former reaction. The precise information about the pion
production cross sections in elementary NN collisions is also 
important for the description of the dynamics of the heavy ion collisions 
within the kinetic theories (eg. BUU) \cite{11,12}.

Although several models have been 
proposed to study the $NN\rightarrow NN\pi$
reactions \cite{13}, only a few have taken the fully relativistic Feynman
diagram approach \cite {14,15,16,17}. However, a systematic study
of all the observables (total, single, double, and triple differential 
cross sections and asymmetries) of $p(p,n\pi^{+})p$ and $p(p,\pi^{0})p$
reactions over a wide range of beam energies within
the covariant model is still lacking,
and the form of several ingredients of the model remains to be far from
being determined unambiguously. Moreover, the role of
 $\sigma$ and $\omega$ mesons have not been explicitly
studied in such theories so far.

Fully relativistic calculations could be necessary even at the beam energies
closer to the $NN \pi$ threshold, where the non-relativistic models
underpredict the experimental data by a factor of approximately 5 
\cite{18}. One
of the consequences of the relativistic effects is that they 
lead to an enhancement of the axial vector current
of the NN system \cite{19}, which can increase the
pion production cross sections, and possibly contribute to an 
explanation of the above discrepancy
(see e.g Horowitz et al. \cite{13}) to a great extent.

The aim of this paper is to perform a detailed investigation of the
$p(p,n\pi^{+})p$ and $p(p,p\pi^{0})p$ reactions using a fully
relativistic Feynman diagram technique. We carry out our calculations
within an effective one-boson exchange (OBE)
nucleon-nucleon (N-N) scattering mechanism, which includes both 
nucleon and delta isobar excitations in the intermediate states.
We consider the exchange of $\pi$, $\sigma$, $\rho$, and $\omega$ mesons.
Most of the parameters of the OBE model are
determined by fitting to the N-N
scattering data \cite{17}. Different delta isobar
propagators proposed in the literature \cite{20,21,22} have been
examined in order to remove the confusion about their most appropriate
form. The relative importance of the contributions of different exchanged
meson in the energy range 300 $MeV$ to 2.0 $GeV$ has been discussed.

In the next section, we give the details of our model and derive the
expressions of various amplitudes. The comparison of our calculations
with the experimental data and the discussion of the results are
presented in section 3. The summary and conclusions of our work are
presented in section 4.

\section{The amplitudes of pion production in One Boson exchange model}
To obtain the desired production cross section for the $NN\rightarrow NN\pi$
processes, we use a fully covariant method based
on an effective one-boson-exchange
model which describes at the same time the elastic nucleon-nucleon scattering.
Pions are simply produced from the external nucleon lines (see Figs.
1a - 1d) or if possible, also from the internal meson lines (Figs. 1e-1h).
The pion production via formation, propagation and the subsequent
decay of the delta isobar is included. It may be noted that in Fig. 1
we have not shown the so called 'pre-emisson' diagrams.
However, in actual calculations their contributions are also included.

\subsection{Model Lagrangian}
The nucleon-nucleon interaction is described by the exchange of
$\pi$, $\sigma$, $\rho$ and $\omega$ mesons in term of which we 
parametrize the T-matrix.
The corresponding Lagrangian densities are given by

\begin{eqnarray}
{\cal L}_{\pi NN} & = & -\frac{f_\pi}{m_\pi} {\bar{\Psi}}_N \gamma _5
                             {\gamma}_{\mu} \tauiso  
                            \cdot (\partial ^\mu {\bf \Phi}) \Psi _N. \\
{\cal L}_{\rho NN} &=&- g_\rho \bar{\Psi}_N \left( \gmu + \frac{k_\rho}
                         {2 m_N} \sigma_{\mu\nu} \partial^\nu\right)
                          \tauiso \cdot \mbox{\boldmath $\rho$}^\mu \Psi_N. \\
{\cal L}_{\omega NN} &=&- g_\omega \bar{\Psi}_N \left( \gmu + \frac{k_\omega}
                         {2 m_N} \sigma_{\mu\nu} \partial^\nu\right)
                          \omega^\mu \Psi_N.   \\
{\cal L}_{\sigma NN} &=& g_\sigma \bar{\Psi}_N \sigma \Psi_N.
\end{eqnarray}

It may be noted that we have used a pseudovector coupling for the $\pi N N$
vertex. In addition the couplings of the mesons to the delta
resonance are needed.
Due to the isospin conservation only  $\rho$ and $\pi$ mesons couple to the
delta resonance. The Lagrangian densities for these processes are

\begin{eqnarray}
{\cal L}_{\pi N\Delta} & = & \frac{f_\pi^\ast}{m_\pi} {\bar{\Psi}}_{\mu}
                            {\bf T} \cdot \partial ^{\mu}
                             {\bf \Phi}_\pi \Psi _N + {\rm h.c.}. \\
{\cal L}_{\rho N\Delta} &=& {\rm i} \frac{g_\rho^{*}}{m_\Delta + m_N}
                        \bar{\Psi}_\mu \mbox{\boldmath $T$} \left(
                        \partial^\nu \mbox{\boldmath $\rho$}^\mu -
                        \partial^\mu \mbox{\boldmath $\rho$}^\nu
                         \right) \gamma_\nu \gf\Psi _N + {\rm h.c.}.
\end{eqnarray}

To insure the gauge invariance of the $\rho - N$ amplitude we must also
include the diagrams 1e - 1h, which depict the processes
where the $\rho$ meson
decays into two pions in flight. The corresponding Lagrangian density is
given by

\begin{eqnarray}
{\cal L}_{\rho \pi \pi} & = & g_{\rho \pi \pi} [(\partial^\mu
                           \mbox{\boldmath $\Phi$}) \times
                     \mbox{\boldmath $\Phi$}] \cdot \rho_\mu.
\end{eqnarray}
In Eqs.( 1-7), $\Psi_N$ represents the Dirac spinor for the
nucleon in the spin-isospin space. $\Psi_\mu$ is the coresponding
Rarita-Schwinger spinor for the delta isobar.
$\mbox{\boldmath $\Phi$}$, $\mbox{\boldmath $\rho$}$, $\sigma$ and $\omega$
represent the pion, rho, sigma and omega meson fields, respectively.
{\bf T} and $\tauiso $ represent the
isospin operator for the transition
$\Delta\rightarrow N\pi$ and the isospin Pauli matrices,  respectively. We
use the conventions and notations of Ref. \cite{23} for definitions of spinors,
operators and isospin matrices, respectively.

Since we use the Lagrangians (1-4) to directly model the T-matrix, we have 
also included a nucleon-nucleon-axial vector-isovector vertex, with the
Lagrangian density given by
\begin{eqnarray}
{\cal L}_{NNA} & = & {\sqrt {g_A}} {\bar {\Psi}} \gamma_5 \gamma_\mu \tauiso \Psi
                     \cdot {\bf {A}}^\mu.
\end{eqnarray}
In Eq. (8) ${\bf A}$ represents the axial vector meson field.
This term is introduced because in the limit of large axial meson masses 
($m_A$) it cures the unphysical behaviour in the angular distribution of 
NN scattering caused by the contact term in the one-pion-exchange 
amplitude \cite{17}, if $g_A$ is chosen to be 
\begin{eqnarray}
g_A =  \frac{1}{3} m_A \left(\frac{f_\pi}{m_\pi}\right)^2 \quad .
\end{eqnarray}
with very large ($\gg m_N$) $m_A$.

The determination of other coupling constants used in this work 
in discussed in the next section.
\subsection{Coupling constants}
The coupling constants appearing in Eqs. (1 - 4) were determined by
fitting to the elastic proton-proton and proton-neutron
scattering data. In the fitting procedure
we also take into account the finite
size of the nucleons by introducing the form-factors

\begin{eqnarray}
F_{i} & = & \left (\frac{\Lambda_i^{2} - m_i^{2}}{\Lambda_i^{2} - q_i^{2}}
        \right ), i= \pi, \rho, \sigma, \omega,
\end{eqnarray}
at each interaction vertex, where q is the four momentum and $m$ the
mass of the exchanged meson. The form factor supresses the contributions
of high momenta and the parameter $\Lambda$, which governs the range of
suppression, can be directly related to the hadron size.
Since the data in the entire range of
beam energies cannot be reproduced with an energy independent set of
parameters, we have used the following energy dependence for the
coupling constants
\begin{eqnarray}
g(\sqrt{s}) & = & g_{0} exp(-\ell \sqrt{s}).
\end{eqnarray}
The parameters ($g_{0}$, $\Lambda$, and $\ell$) were determined \cite{17} by
fitting to the relevant proton-proton and proton-neutron
data at three beam energies of 1.73 $GeV$, 2.24 $GeV$,
and 3.18 $GeV$, where a good fit to the elastic scattering amplitudes
were obtained. Table 1 shows the values of the parameters obtained by
this procedure. It may be noted that for the case of pion we have shown
in this table the constant $g_\pi$ which is related
to $f_\pi$ of Eq. (1) as $g_\pi \; = \;  (f_\pi/m_\pi) 2m_N$.

The value of the coupling constant $f_{\pi}^{*}$ for the $\pi N\Delta$
vertex has been determined from the
$\Delta \rightarrow N + \pi $ decay and its
value is 2.13. As in the case of NN scattering,
we use form factors for the
$N\, - \,\Delta$ vertices as well, which have a dipole form \cite{16,17}
\begin{eqnarray}
F_{i}^{*} & = & \left ( \frac{\Lambda_i^{*2} - m_i^{2}}{\Lambda_i^{*2} - q_i^{2}}
                \right )^2 \quad, i= \pi, \rho.
\end{eqnarray}
The parameters $g_{\rho}^{*}$, $\Lambda_{\pi}^{*}$ and $\Lambda_{\rho}^{*}$
were determined by fitting to mass differential cross sections for the
reaction $N + N \rightarrow N + \Delta$ in the energy range of 1-2.5 $GeV$.
Their values are,
\begin{eqnarray}
\Lambda_{\pi}^{*} & = & 1.421 \, GeV, \nonumber \\ 
\Lambda_{\rho}^{*} & = & 2.273 \, GeV,\nonumber \\
g_{\rho}^{*} & = & 7.4.
\end{eqnarray}
For the graphs 1e - 1h, we have taken
$g_{\rho \pi \pi}\, = \, 2g_\rho$ \cite{24,25}.
In this way the $NN\,meson$ and $N \Delta$ vertices are determined rather
reliably. We then assume that their off-shell dependence is detemined 
solely by the form factors (10) and (12 ).

For the delta isobar propagator we used the form given by Benmerrouche
et al. \cite{22}
\begin{eqnarray}
G_{\mu \nu}^\Delta (p) & = & -\frac{i(p\!\!\!/ + m_\Delta)}{p^2 - m_\Delta ^2}\nonumber \\
                       &   & [g_{\mu \nu} - \frac{1}{3}\gamma_\mu \gamma_\nu 
                              - \frac{2}{3m_\Delta^2} p_\mu p_\nu
                              + \frac{1}{3m_\Delta^2}
                                ( p_\mu \gamma_\nu - p_\nu \gamma_\mu )].
\end{eqnarray}
Similar expression for the $\Delta$ propagator has been derived 
also in Ref. \cite{26}. It should be noted that the form of the
propagator for an interacting delta isobar remains the same
as in Eq. (14) \cite{22}. In appendix B, we have presented a review 
of the derivation of this propagator and have discussed 
the difficulties associated with $\Delta$ isobar propagators presented
by Williams and other authors \cite{20,21}.

The mass of the delta isobar $m_\Delta$ appearing in the denominator
term $(p^2 - m_\Delta^2)$ is modified by adding to it an imaginary width,
which is a function of the pion-nucleon center of mass momentum
${\bf{p}}_\pi^\prime$. This is due to the fact
that $\Delta$ isobar is not a stable particle and it decays with a width
which varies with its mass \cite{27}. There are
several prescriptions for the mass dependent delta isobar width given
in the literature \cite{16,28,29}.
We have performed calculations by using the following forms
\begin{eqnarray}
\Gamma_D(p_\pi^\prime) & = & \Gamma_0
                    \left ( \frac{p_\pi^{\prime 3}}{p_\pi^{R 3}} \right )
 \left ( \frac{p_\pi^{R 2} + \epsilon^2}{p_\pi^{\prime 2} + \epsilon^2} \right ),
\end{eqnarray}
where
\begin{eqnarray}
p_\pi^{\prime 2} & = & \frac{[p_i^2 -(m_N - m_\pi)^2][p_i^2 - (m_N + m_\pi)^2]}
                            {4p_i^2}.
\end{eqnarray}
In Eq. (16), $p_i$ is the four-momentum of the
intermediate delta isobar (see Fig. 1).
$p_\pi^R$ used in Eq. (15) is obtained from Eq. (16) by substituting
$p_i^2 \; = \; m_\Delta^2$. The constant $\Gamma_0$ is the free delta
width; its value is taken to be 0.120 $GeV$ and $\epsilon \; = \; 0.16$.
This form has been used by Dmitriev et al. \cite{28}. Verwest \cite{15} has 
used a somewhat different form which is given by 
\begin{eqnarray}
\Gamma_{ver}(p_\pi^\prime) & = & \Gamma_0
           \left ( \frac{p_\pi^{\prime 3}}{p_\pi^{R 3}} \right )
      \left ( \frac{\sqrt{p_\pi^{\prime 2} + m_\pi^2} + m_N}
            {2m_N} \right ),
\end{eqnarray}
where $\Gamma_0$ has the same value as given above.

In the next subsection we present the expressions for the
amplitudes of various graphs as shown in Fig. 1.

\subsection{Amplitudes and cross sections}

After having established the interaction Lagrangians and the form
of the delta isobar propagator, we can now proceed to calculate the
amplitudes corresponding to the various diagrams associated with the reactions
$p(p,n\pi^{+})p$ and $p(p,p\pi^{0})p$. The Feynman rules for writing
down these amplitudes are well known (see e.g. \cite{30}). The isospin part
is treated separately which gives rise to a constant factor for each graph.
In table 2 we give the values of these factors for the
post-emission (shown in Fig. 1) as well as for the pre-emission graphs
corresponding to the emission of $\pi^{+}$ and $\pi^{0}$ mesons. Values
for the pion emission from the intermediate states (Figs. 1e-1h)
are shown in table 3. It may be noted that in these tables the values for
the $\pi^+$ case are calculated by assuming particles 3 and 4 as neutron
and proton respectively. Isovector corresponds to the exchange of $\pi$
and $\rho$ mesons while isoscalar to that of $\sigma$ and $\omega$ mesons.

In the following we give the expressions for the amplitudes for the
pion production processes via excitation of both nucleon
and delta isobar intermediate states for one graph eg. Fig. 1c. We
also give expression for one diagramm (Fig. 1f) where pions are emitted
from an intermediate meson.
The amplitudes for other similar graphs
can be written in a straight forward manner.

{\bf {(i) Nucleon intermediate state and pion exchange}}

\begin{eqnarray}
 A_N^{\pi} (c) & = & - Q_{N}^{\pi}(c) \left( \frac{f}{m_\pi} \right)^3
              {\bar {\psi}}(p_3)\gamma_5 \gamma_\lambda q^\lambda
              \psi(p_1) D_\pi (q) \nonumber \\
        &   & \times {\bar {\psi}}(p_4) \gamma_5 \gamma_\mu p_\pi ^\mu
              D_N(p_i) \gamma_5 \gamma_\nu q^\nu \psi(p_2),
\end{eqnarray}
where $D_\pi(q)$ and $D_{N}(p_i)$ are the propagators for the exchanged
pion and the intermediate nucleon respectively, which are defined as
\begin{eqnarray}
 D_\pi (q) & = & \frac{i}{q^2 - m_\pi ^2}, \\
 D_{N} (p_i) & = & i \frac{p_{i\eta} \gamma^\eta + m_N}{p_i^2 - m_N^2}.
\end{eqnarray}
Various momenta appearing in Eq. (18) are defined in Fig. 1. The intermediate
momenta are given by 
$q = p_1$ - $p_3$ and $p_i=p_\pi$ + $p_4$. $\psi$ is the Dirac spinor
in the spin space and $Q_N^{\pi}(c)$ is the isospin coupling factor for the
nucleon pole as shown in Table 2.

{\bf{(ii) Delta isobar intermediate state and pion exchange}}
\begin{eqnarray}
 A_\Delta^{\pi} (c) & = & - Q_{\Delta}^{\pi}(c) \left( \frac{f}{m_\pi} \right)
               \left( \frac{f^*}{m_\pi} \right)^2
              {\bar {\psi}}(p_3)\gamma_5 \gamma_\lambda q^\lambda
              \psi(p_1) D_\pi (q) \nonumber \\
        &   & \times {\bar {\psi}}(p_4) p_\pi ^\mu
              G_{\nu \mu}^\Delta(p_i) q^\nu \psi(p_2),
\end{eqnarray}
where $G_{\mu \nu}^\Delta(p_i)$ is the delta isobar propagator
as discussed in the previous subsection, and $Q_\Delta^{\pi}(c)$ is the
isospin coupling factor for the delta pole as shown in Table 2.

{\bf{(iii) Nucleon intermediate state and $\rho$ meson exchange}}
\begin{eqnarray}
 A_N^{\rho} (c) & = & - Q_{N}^{\rho}(c) \left( \frac{f}{m_\pi} \right)
                   g_\rho^2
              {\bar {\psi}}(p_3) ( \gamma_\mu + \frac{ik_{\rho}}{2m_N}
              \sigma_{\mu \nu} q^\mu)
              \psi(p_1) D_\rho^{\nu \alpha} (q) \nonumber \\
        &   & \times {\bar {\psi}}(p_4) \gamma_5 \gamma_\lambda p_\pi ^\lambda
              D_N(p_i)( \gamma_\alpha - \frac{ik_{\rho}}{2m_N}
               \sigma_{\alpha \beta} q^\beta) \psi(p_2),
\end{eqnarray}
where $D_\rho (q)$ is the propagator for the $\rho$ meson defined as
\begin{eqnarray}
D_\rho (q)^{\mu \nu} & = & -i \left( \frac{g^{\mu \nu} - \frac{q^\mu q^\nu}{q^2}}
                            {q^2 - m_\rho ^2} \right ),
\end{eqnarray}
with $g^{\mu \nu}$ being the usual metric tensor. In Eq. (22)
$\sigma_{\mu \nu}$ is defined as
\begin{eqnarray}
\sigma_{\mu \nu} & = & \frac{i}{2}(\gamma_\mu \gamma_\nu - \gamma_\nu \gamma_\mu),
\end{eqnarray}
Other quantities remain the same as earlier.

{\bf{(iv) Delta isobar intermediate state and $\rho$ meson exchange}}
\begin{eqnarray}
 A_\Delta^{\rho} (c) & = & - Q_{\Delta}^{\rho}(c) \left( \frac{f^*}{m_\pi}
                             \right) g_\rho \frac{g_\rho^*}{m_\Delta + m_N}
              {\bar {\psi}}(p_3) ( \gamma_\alpha + \frac{ik_{\rho}}{2m_N}
              \sigma_{\alpha \nu} q^\nu)
              \psi(p_1) \nonumber \\
        &   & \times D_\rho^{\alpha \nu} (q) 
              {\bar {\psi}}(p_4) p_\pi ^s G_{ts}^\Delta(p_i)
              (q^\mu g^{t \nu} - q^t g^{\mu \nu})\gamma_\mu \gamma_5 \psi(p_2),
\end{eqnarray}
In Eq. (25) all the quantities are the same as defined earlier.

{\bf{(v) Nucleon intermediate state and $\sigma$ meson exchange}}
\begin{eqnarray}
A_N^{\sigma} (c) & = & -Q_{N}^{\sigma} g_{\sigma}^2 \left (\frac{f_\pi}{m_\pi} \right)
  {\bar {\psi}}(p_3) \psi(p_1)  D_\sigma(q) {\bar {\psi}}(p_4) \nonumber \\
                 &   & \times \gamma_5
                     \gamma_\mu p_{\pi}^\mu D_N(p_i) \psi(p_2),
\end{eqnarray}
where $D_\sigma (q)$ is the propagator for the sigma meson whose form
is the same as that given by Eq. (19) except that
in the denominator the pion mass is replaced by that of the sigma meson.

{\bf{(vi) Nucleon intermediate state and  $\omega$ meson exchange}}

In this case the form of the amplitude is the same as that of the
$\rho$ meson exchange.

{\bf{(vii) Pion emission from the decay of $\rho$ meson in the
intermediate states}} (Fig.1f)

\begin{eqnarray}
A^{\rho \pi \pi} (f) & = & -ig_\rho g_{\rho \pi \pi} \frac{f_\pi}{m_\pi}
                  {\bar{\psi}}(p_3)(\gamma_\mu + \frac{ik_\rho}{2m_N}
                  \sigma_{\mu \nu} q^\nu) \psi (p_1) D_\rho^{\nu \mu}
                  p_{\pi \nu} \nonumber \\
                    &   &  \times D_\pi(q^\prime) {\bar {\psi}}(p_4) \gamma_5 \gamma_\eta
                  q^{\prime \eta} \psi (p_2),
\end{eqnarray}
where $q^\prime\, = \, p_4\, - \, p_2$.

{\bf{(viii) Nucleon intermediate state and heavy axial meson exchange}}

\begin{eqnarray}
 A_N^{A} (c) & = & - Q_{N}^{A}(c) g_{A} \left( \frac{f_\pi}{m_\pi} \right)
              {\bar {\psi}}(p_3)\gamma_5 \gamma_\mu \psi(p_1)
               D_{A}^{\mu \nu} (q) \nonumber \\
        &   & \times {\bar {\psi}}(p_4) \gamma_5 \gamma_\mu p_\pi ^\mu
              D_N(p_i) \gamma_5 \gamma_\nu \psi(p_2),
\end{eqnarray}
where $D_{A}^{\mu \nu}$ is the propagator for the axial vector meson
which is defined by
\begin{eqnarray}
D_{A}^{\mu \nu}(q) & = & -i \left( \frac{g^{\mu \nu}}
                            {q^2 - m_{A}^2} \right )
\end{eqnarray}

In Eq. (29), the mass of the axial meson was taken to be very large
(188 $GeV$), as the corresponding amplitude is that of the contact
term.
 
The amplitudes given above can be simplified by contracting out the gamma
matrices using the Dirac equation whenever applicable. The simplified
expressions are given in Appendix A.
The total amplitude is obtained by summing (coherently) the amplitudes
corresponding to all the graphs. It should be noted that the
exchange graphs have an extra minus sign due the antisymmetrisation.

The general formula for the invariant cross sections of the
$ N\,\,+ \,\,N\,\,=\,\,N\,\,+\,\,N\,\,+\,\,\pi$ process is written as
\begin{eqnarray}
d \sigma & = & \frac{m_N^4}{2 \sqrt{(p_1 \cdot p_2)^2 - m_N^4}}
               \frac{1}{(2\pi)^5}
               \delta^4(P_f - P_i)  | A_{fi} |^2 \Pi_{a=1}^3
               \frac{d^3{\bf{p}}_a}{E_a},
\end{eqnarray}
where $A_{fi}$ represents the sum of all the amplitudes, $P_i$ and
$P_f$ the sum of all the momenta in the intial and final states respectively,
and $p_a$ the momenta of the three particles in the final state.
The corresponding cross sections in the laboratory or center mass systems
can be written from Eq. (30) by imposing the relevant conditions.
\section{Comparison to data and discussions}

The model presented in the previous section has been used to study the
available data on the total, double and triple differential cross
sections for the
$p + p \rightarrow n + \pi^{+} + p$ and
$p + p \rightarrow p + \pi^{0} + p$ reactions for beam energies
300 MeV to 2 GeV. We emphasis here that the parameters
described in the previous section have been kept fixed
throughout in all the calculations described
subsequently.

\subsection{Total pion production cross section}

In Figs. 2a and 2b, we show a comparison of our calculations with
the experimental total cross sections for the reactions
 $p \; + \; p \rightarrow p \; + \; n \; + \; \pi^{+}$ and
 $p \; + \; p \rightarrow p \; + \; p \; + \; \pi^{0}$ respectively,
as a function of beam energy. The dashed lines represent the results
obtained by including only those graphs where pion production
proceeds via intermediate delta isobar states, while the
dashed-dotted  
lines give the results with only nucleon intermediate states.
The solid lines shows the results where all the graphs are
included. We note that the measured cross sections are
reproduced reasonably well by our calculations
in the entire range of beam energies. This is
remarkable in view of the fact that none of the parameters of the model
have been adjusted to the pion production data. Furthermore,
they were determined by fitting the
NN elastic scattering data (differential cross sections)
at beam energies above 1 GeV. The same
set of parameters (with an energy dependence of the coupling
constants given by Eq.  (11) seems to reproduce the data
at low energies as well.

 The contribution of the delta isobar excitation is not
important at beam energies below approximately 350 $MeV$, which is due
to the fact that at lower energies pions are predominantly in a relative
$S$-state; thus the possibility of forming a delta isobar is greatly
reduced. However, the pion production is dominated by the $\Delta$
isobar excitation at higher beam energies.

In Fig. 3, we show the contribution of various meson exchange processes
to the total cross section of the $(p,n\pi^{+})$ reaction in the considered
range of beam energies. The contributions
of the heavy axial meson exchange are not shown in this figure 
as they are negligibly small. 
We note that the pion exchange graphs dominate the
production process for all the energies.
The contribution of the $\rho$ meson exchange 
is almost negligible at lower beam energies, and even in the energy
range of 1 - 2 $GeV$, it is at least an order of magnitude smaller than
that of the pion exchange process. Dmitriev et al. \cite{28}
have made the similar observation for the reaction
$p \; + \; p \rightarrow n \; + \; \Delta^{++}$ 
within a similar type of model. This lends a {\it posteriori}
credence to the calculations presented in Ref. \cite{10}
for the $A(p, \pi)B$ reaction
within a covariant two-nucleon model at a beam energy
of 800 $MeV$, where only one-pion exchange intermediate
processes were considered.

At the incident energies very close to the pion
production threshold the contributions of the $\sigma$
and $\omega$ meson exchange are as strong as
that of the pion exchange.
Therefore calculations, which do not include these mesons,
underestimate the pion production cross sections near the
threshold (see eg. Horowitz et al. in [13], and \cite{31}).

As we note from Table 1, the cut-off parameters $\Lambda$
must have values in the range of 1.0 - 1.6 $GeV$ in order
to fit the NN scattering data. Similar values
for this parameter have been used in almost all the earlier
calculations\cite{10,32,33,34} of the proton
induced pion production (which mostly included
only pion exchange contributions, requiring therefore
only $\Lambda_\pi$ and $\Lambda_\Delta$ which were
taken to be the same). In Fig. 4 we show the total
cross sections for the $p(p,n \pi^{+})p$ reaction
as a function of beam energy for two values of $\Lambda_\pi$.
The solid and the long dashed lines show the results obtained
with values 1.005 $GeV$ and 0.63 $GeV$ respectively for this
parameter. It is apparent that the latter value of
$\Lambda_\pi$ leads to
a smaller cross section and a poorer fit to the data.
Smaller values of $\Lambda$ imply a reduction in the range
of the pion exchange contribution to the NN interactions which
reduces its contribution to the pion production cross section as
this is the dominant process  as has been discussed earlier.
Although a value of $\Lambda_\pi \leq \; 0.8$
is consistent with the chiral bag model \cite{35},
the quantitative description of the NN data requires
a value $\geq$ 1.0 $GeV$ \cite{36}. We therefore,
stick to the value of $\Lambda_\pi$ as shown in Table 1.

In Fig. 5, we investigate the effect of using the pseudoscalar
coupling for the $\pi N N $ vertex on the $\pi^{+}$ production
cross section. In this figure we show the ratio of the
total cross sections obtained by using the pseudovector
($\sigma_{P V}$) and pseudoscalar ($\sigma_{P S}$) couplings
for the $\pi N N $ vertex as a function of beam energy.
The solid line represents the results obtained by including
all the graphs while the long dashed line give the results
obtained with only nucleon intermediate excitations.
It is clear that the pion production is enhanced if the 
pseudoscalar coupling is used. This effect is very large at
lower beam energies. At higher beam energies, where
$\Delta$ isobar excitation dominates, this effect still persists
even when all the graphs are included in the calculations.

In the one pion exchange picture of the NN interaction, when
both nucleon are on the mass shell, it will not be possible
to make a difference between the pseudovector and
pseudoscalar couplings, as both will produce the same
result. However, in Fig. 1, one of the nucleons can go
off-shell. Therefore, in this case the two couplings
will produce different results. One can show (taking the
case of just one diagram, say for example Fig. 1a) that
the ratio of the amplitudes obtained by using pseudovector
and psedoscalar couplings is given approximately by
$(E_\pi/2m_N)$. At lower beam energies this ratio is very small,
which explains to some the extent the trend seen in Fig. 5.

In general, the pseudovector coupling for the $\pi N N$ vertex
is preferred. The $p$-wave interaction between pions and nucleons
is very strong and attractive; the formation of the $\Delta$ isobar
is the consequence of this interaction. On the other hand,
the $s$-wave interaction is very weak. The pseudovector
coupling automatically
incorporates these features due to the derivative term.
One can show purely on formal grounds that it is also
consistent with the
constraint imposed by the PCAC hypothesis \cite{19}.
The derivative of the
axial vector current operator based on the pseudovector $\pi NN$
coupling vanishes as the pion mass goes to zero. The pseudoscalar
coupling does not satisfy this criterion; it also produces a $s$-wave
interaction which is much too strong \cite{37}. Furthermore,
this coupling has a history of problems in
Dirac calculations \cite {38}.
Therefore we do not consider the pseudoscalar couplings
for the $\pi NN$ vertex in our calculations.

\subsection{Double differential cross section}

We have also calculated the double differential cross sections
for the $pp \rightarrow \pi^{+} X$ reactions and have
compared the results with one set of existing experimental
data \cite{39} at the proton incident energy of
800 $MeV$, in Fig. 6; this set of data were also
analysed by Verwest \cite{15}. It is clear that
the contributions of the delta isobar excitation (dashed lines)
dominate the total cross sections (solid lines),
whereas those of the nucleon intermediate state
(dotted lines) are very small. The
agreement between theory and the data is reasonably good
at forward pion angles. However, at higher pion angles our
calculations overpredict the data by factors of 1.5 to 2. The peak 
cross sections are affected by the different multipolarities (dipole or 
monopole) of the form-factors (Eq. (10) - (11)) and forms of the 
delta decay width $\Gamma_D$ (Eq. (15)). For instance,
with a monopole form factor for the the $N \Delta \rho$ vertex,
the cross sections near the peak are reduced by approximately 20 $\%$. 
On the other hand, using the
width $\Gamma_{ver}$ enhances the cross sections
near the peak region further by about 5-6 $\%$. 
There is, therefore, some scope to explain the small disagreement
between data and out calculations at larger pion angles. This also  
indicates that these measurments
may be useful in differentiating between various forms of the 
formfactors and delta decay widths.
 
The experimental
points towards the larger momentum ends of the spectra come from the
processes like $\pi d$ final states which is obviously not included in
our calculations.

\subsection{Triple differential cross section}

We now compare our results with data taken in kinematically complete
measurments where all the three particles in the final channels are
measured. In Fig. 7, we present a comparison of our calculations
with the data \cite{5} for
the triple differential cross sections for two sets of pion and
proton angles as a function of outgoing proton momentum at the
beam energy of 800 $MeV$. The peak reagion in this figure corresponds
to the final state proton momenta associated with the formation of the
delta isobar and its decay. The solid lines (dashed lines)  
in this figure show the results of calculations which include
the contributions of all the graphs (only delta isobar intermediate
states). Clearly the peak region is dominated by the delta excitation 
process. 
At higher angles and in the regions away from the delta excitation
peak, the contributions of nucleon excitation are not negligible as
can be seen in the lower part of this figure. The shift in the peak
position towards higher momenta at larger angles is due to the
fact that at larger pion angles the delta isobar formation is
associated with larger outgoing pion momenta.

 It should, however, be 
noted that the position of the peak in the calculated
cross sections is shifted towards somewhat larger momenta
as compared to the experimental data. Obviously the form of
the delta decay width will play a crucial role
in determining the peak position. In Fig. 7 we have used 
the decay width $\Gamma_{ver}$.
The effect of using other forms of the deacay width
is shown in Fig. 8, where the solid lines represent the results of
the calculations performed with the delta decay width of Verwest,
while the dashed lines corresponds to that of Dmitriev.
The short dashed lines
show the results obtained with Dmitriev's form but with a value of
$\Gamma_{0} \; = \; 0.100 \; GeV$ \cite{5}. We see that the
peak positions of the experimental cross sections are
correctly reproduced by calculations using the Decay width of
Dmitriev. Hoewever, the absolute magnitudes of the cross sections
are a bit too high at larger pion angles.

\section{Summary and conclusions}

In this paper we have presented a fully covariant one-boson
exchange model to describe the $p(p,n\pi^{+})p$ and
$p(p,p\pi^{0})p$ reactions in the beam energy ($E_{beam}$)
range of 300 MeV to 2 GeV. The model contains
only the physical parameters (like coupling constants and
cutoff masses), which were determined by fiting to the NN
elastic scattering data at three beam energies above 1 $GeV$.
All the parameters so determined were
held fixed throughout the considered energy range; none of
the parameters were adjusted to the pion production
data of any kind. We considered the exchange of the
$\pi$, $\rho$, $\sigma$ and $\omega$ mesons; the latter
two were not considered in earlier
calculations of the $NN \pi$ processes. The calculations
included pre- and post-emission graphs and considered
the excitation of both the delta isobar and nucleon intermediate
states. The deacy of the $\rho$
meson in flight (the so-called intermediate emission diagram)
was also incorporated in our calculations.

We found that the pion exchange processes dominate the cross sections
in the entire energy region. The contribution of the $\rho$ meson
exchange is almost negligible at lower beam energies. Even in the
energy range of 1 - 2 $GeV$ its contribution is less than 10 $\%$. This is
an important observation as it implies that in the
description of the $A(p,\pi)B$ reaction at beam energies around 1 $GeV$
within the covariant two nucleon
model, it suffices to consider only pion exchange intermediate
states, which reduces the intricacy of the calculations
quite a bit \cite{10}. The exchange of $\sigma$ and $\omega$ mesons is
important for beam energies closer to the pion production threshold,
while at
intermediate and higher beam energies their contributions are of the
same order of magnitude as that of the $\rho$ meson exchange. Therefore,
it would be important to include the exchange of both $\sigma$ and
$\omega$ mesons in order to explain the near threshold
data on pion production in $pp$
collisions taken recently at the Bloomington cooler
cyclotron \cite{8}. It should be noted that a fully covariant
calculation is necessary at even  
these energies because the non-relativistic reduction
of the $\pi NN$ Lagrangian has ambiguities, which cast doubt
on  the results obtained within such approaches\cite{10,40}.

The excitation and decay of the delta isobar dominates the pion
production processes at $E_{beam} \geq 0.5 \; GeV$ where it accounts
for almost entire cross sections on its own.
However, this process is not
important for lower beam energies; it contributes almost negligibly
to pion production near the threshold.

We have also reviewed the derivation of the covariant delta
isobar propagator and have showed
that the propagator given by Williams does not
represent the correct propagator for a massive spin-$3/2$ field.
This point was further stressed by performing numerical calculations
with this propagator. We have found that the Williams propagator leads to
very large cross sections ( which look almost like a pole ) for total
as well as differential cross sections in certain energy regions which is
clearly unphysical.

The calculations performed with a pseudoscalar coupling for the
$\pi NN$ vertex were found to produce too large cross sections
particularly near the pion production threshold. This is due to
the fact that pseudoscalar coupling alone (without any $\sigma \pi$
coupling) produces a much too large $S$-wave
$\pi N$ interaction, which dominates the pion production
near the threshold. However, this is clearly not in agreement with the
experimental data.
The pseudoscalar coupling is not consistent with
the constraints of the PCAC. Therefore, we have preferred to use the
pseudovector couplings in our calculations.

We have also analysed the available double
and triple differential cross sections for pion production and found
that our model provides a reasonable description of these data as well.
The triple differential cross sections are sensitive to the form of the
momentum dependent delta decay widths and probably also to the 
form of the cut-off parameters. As there are
many versions of this width available in the literature, the complete
kinematical detection of three particles in the final channel
may perhaps help in making distinction between them. This would lead to
a better understanding of the off-shell behaviour of the delta isobar
\cite{41}.

In this work we established a covariant framework to describe the
inelastic channels of the NN collisions. and used this to investigated 
the strongest of them, the pion emission. This should
now form the basis for the study of relatively weaker inelastic processes
like emission of $\eta$ mesons and also the $\eta^\prime$ meson which
is thought to provide a novel tool to study the baryonic resonances
around 2 $GeV$ \cite{42}

\newpage
\begin{appendix}
\section*{Appendix A}
By using the complicated but straight-forward algebra of
$\gamma$ matrices and the Dirac equation for a free particle
the amplitudes given in subsection (2.3) can be rewritten into 
relatively simple forms which are suitable for numerical
calculations. We have dropped the isospin factors in the expressions
shown below, however in actual calculations they are included.

{\bf{(i) Nucleon intermediate state and the pion exchange}}
\begin{eqnarray}
 A_N^{\pi} (c) & = &- \left( \frac{f}{m_\pi} \right)^3
                     \frac{1}{q^2 - m_\pi^2} \frac{1}{p_i^2 - m_N^2}
                     {\bar {\psi}}(p_3)\gamma_5 \psi(p_1) \nonumber \\
               &   & \times  {\bar {\psi}}(p_4)
               (E + F \gamma_\mu p_\pi^\mu) \psi(p_2),
\end{eqnarray}
where
\begin{eqnarray}
E & = & 2 m_N (m_N^2 - p_i^2) \\
F & = & p_i^2 + 3 m_N^2.
\end{eqnarray}
It may be noted that if the pseudoscalar coupling for the $ \pi N N $
vertex is used, we get the same expression for this amplitude but the
constants $E$ and $F$ are defined in the following way
\begin{eqnarray}
E & = & 0 \\
F & = & 4 m_N^2.
\end{eqnarray}

{\bf{(ii) Delta isobar intermediate state and the pion exchange}}
\begin{eqnarray}
 A_\Delta^{\pi} (c) & = & - 2m_N \left( \frac{f}{m_\pi} \right )
                            \left( \frac{f^*}{m_\pi} \right)^2
                            \frac{1}{q^2 - m_\pi^2}
                            \frac{1}{p_i^2 - m_\Delta^2} \nonumber \\
                    &   &  \times {\bar {\psi}}(p_3) \gamma_5 \psi(p_1)
                           {\bar {\psi}}(p_4)
                           (G + H \gamma_\mu p_\pi^\mu) \psi(p_2).
\end{eqnarray}
In Eq. (36) the expressions of $G$ and $H$ will depend on the form of the
delta isobar propagator used in the derivations. For the propagator
$G_{\mu \nu}^\Delta$ (Eq. (14)), one gets
\newpage
\begin{eqnarray}
G & = & m_N qp_\pi - \frac{2}{3} m_N p_i p_\pi
                   + \frac{1}{3} m_N m_\pi^2
                   - \frac{2p_i p_\pi p_i q}{3m_\Delta^2} m_N
                   + \frac{2(p p_\pi)^2}{3 m_\Delta} \nonumber \\
  &   &            - \frac{p_i p_\pi m_\pi^2}{3 m_\Delta}
                   + \frac{p_i q m_\pi^2}{3 m_\Delta}
                   + m_\Delta q p_\pi
                   - \frac{2}{3} m_\Delta p_i p_\pi
                   + \frac{1}{3}  m_\Delta m_\pi^2 \nonumber \\
  &   &            - \frac{2 p_i p_\pi p_i q}{3 m_\Delta}, \\
H & = & qp_\pi - \frac{2}{3} m_N^2
               + \frac{1}{3} m_\pi^2
               - \frac{2p_i p_\pi p_i q}{3m_\Delta^2}
               - \frac{p p_\pi m_N}{3 m_\Delta} \nonumber \\
  &   &        - \frac{p_i q m_N}{3 m_\Delta}
               + \frac{2}{3} m_\Delta m_N
               + \frac{1}{3} p_i p_\pi
               - \frac{1}{3} p_i q.
\end{eqnarray}
Whereas with the propagator $G_{\mu \nu}^{\Delta W}$
(see appendix B), we obtain
\begin{eqnarray}
G & = & m_N qp_\pi - \frac{2}{3} m_N p_i p_\pi
                   + \frac{1}{3} m_N m_\pi^2
                   - \frac{2p_i p_\pi p_i q}{3p_i^2} m_N
                   + \frac{2(p p_\pi)^2}{3 p_i^2} m_\Delta \nonumber \\
  &   &            - \frac{p_i p_\pi m_\pi^2}{3 p_i^2} m_\Delta
                   +  \frac{p_i q m_\pi^2}{3 p_i^2} m_\Delta
                   + m_\Delta q p_\pi - \frac{2}{3} m_\Delta p_i p_\pi
                   + \frac{1}{3} m_\Delta m_\pi^2 \nonumber \\
  &   &            - \frac{2 p_i p_\pi p_i q}{3 p_i^2} m_\Delta, \\
H & = & qp_\pi - \frac{2}{3} m_N^2
               + \frac{1}{3} m_\pi^2
               - \frac{2p_i p_\pi p_i q}{3 p_i^2}
               - \frac{p p_\pi m_N}{3 p_i^2} m_\Delta \nonumber \\
  &   &        - \frac{p_i q m_N}{3p_i^2} m_\Delta
               + \frac{2}{3} m_\Delta m_N + \frac{1}{3} p_i p_\pi
               - \frac{1}{3} p_i q.
\end{eqnarray}

{\bf{(iii) Nucleon excitation and $\rho$ meson exchange}}

We make use of the following relations
\begin{eqnarray}
q_\mu \left ( g^{\mu \nu} - \frac{q^\mu q^\nu}{q^2} \right ) & = & 0, \\
q^\mu \sigma_{\mu \nu} q^\nu & = & 0, \\
{\bar {\psi}}(p_3) \gamma_\mu \psi(p_1)(p_3^\mu - p_1^\mu) & = & 0,
\end{eqnarray}
to reduce the amplitude given by Eq. (22) as follows
\begin{equation}
A_N^\rho (c) = g_\rho^2 \left (\frac{f}{m_\pi} \right )
               \frac{1}{q^2 - m_\rho^2} \frac{1}{p_1^2 - m_N^2}
               \psi(p_4)(D_1 - D_2)\psi(p_1).
\end{equation}
In Eq. (44) $D_1$ and $D_2$ are defined as
\begin{eqnarray}
D_1 & = & (\gamma_\mu p_\pi^\mu \gamma_\nu p_i^\nu \gamma_\eta b^\eta
         - m_N \gamma_\mu p_\pi^\mu \gamma_\eta b^\eta) \gamma_5, \\
D_2 & = &  - \left (\frac{k_\rho}{4 m_N} \right )
          [ -m_\pi^2 \gamma_\mu q^\mu \gamma_\nu b^\nu \gamma_5
           - m_N \gamma_\mu p_\pi^\mu \gamma_\nu
            q^\nu \gamma_\eta b^\eta \gamma_5 \nonumber \\
    &   &  + 2 p_i p_\pi \gamma_\mu q^\mu \gamma_\nu b^\nu \gamma_5
           -m_N \gamma_\mu p_\pi^\mu \gamma_\nu
            q^\nu \gamma_\eta b^\eta \gamma_5 \nonumber \\
    &   &  + m_\pi^2 \gamma_\mu b^\mu \gamma_\nu q^\nu \gamma_5
           + m_N \gamma_\mu p_\pi^\mu \gamma_\nu
             b^\nu \gamma_\eta q^\eta \gamma_5 \nonumber \\
    &   &  -2 p_i p_\pi \gamma_\mu b^\mu \gamma_\nu q^\nu \gamma_5
           + m_N \gamma_\mu p_\pi^\mu \gamma_\nu
            b^\nu \gamma_\eta q^\eta \gamma_5],
\end{eqnarray}
where
\begin{equation}
b_\mu = {\bar {\psi}}(p_3) [ (1-k_\rho) \gamma_\mu + \frac{k_\rho}{m_N}
                    (p_{3 \mu } - q p_3 q_\mu) \psi(p_1).
\end{equation}
{\bf{ (iv) Delta isobar excitation and $\rho$ meson exchange}}

Using Eqs. (41) - (43) we can write the amplitude given by
Eq. (25) as 
\begin{eqnarray}
A_\Delta^\rho (c) & = & -g_\rho \frac{g_\rho}{m_\Delta + m_N} \left (
                       \frac{f_\pi^*}{m_\pi} \right ) {\bar {\psi}}(p_3)
  [(1 - k_\rho) \gamma^\nu + \frac{k_\rho}{m_N}p_3^\nu] \psi(p_1) \nonumber \\
                   &   & \times \psi(p_4)
                  ( p_\pi^s G_{st}^\Delta \gamma_\mu q^\mu \gamma_5
                  -p_\pi^s G_{st}^\Delta q^t \gamma_\nu \gamma_5 ).
\end{eqnarray}
This equation can be further simplified by using the forms of
the propagator $G_{st}^\Delta$ as discussed earlier. These expressions
are not being given here as they are very lengthy even though
it is straightforward to derive them.

{\bf{(v) Nucleon excitation and $\sigma$ meson exchange}}
\begin{eqnarray}
A_N^\sigma (c) & = & \left ( \frac{f_\pi}{m_\pi} \right ) g_\sigma^2
   \frac{1}{q^2 - m_\pi^2} \frac{1}{p_i^2 - m_N^2}
                    {\bar {\psi}}(p_3) \psi(p_1) \nonumber \\
              &   & \times {\bar {\psi}}(p_4)
               (2m_N \gamma_\mu p_\pi^\mu + m_\pi^2 - 2 p_i p_\pi)
               \gamma_5 \psi(p_2).
\end{eqnarray}

{\bf{(vi) Pion emission from the decay of $\rho$ meson in the
intermediate states}}
\begin{eqnarray}
A^{\rho \pi \pi} & = & i 2m_N g_\rho g_{\rho \pi \pi}
          \left ( \frac{f_\pi}{m_\pi} \right ) \frac{1}{q^2 - m_\rho^2}
                   \frac{1}{q^{\prime 2} - m_\pi^2} \nonumber \\
          &   & \times {\bar {\psi}}(p_3)[(1 - k_\rho) \gamma_\mu p_\pi^\mu
+ \frac{k_\rho}{m_N}(p_3 p_\pi - \frac{p_3 q}{q^2} q p_\pi)] \psi(p_1) \nonumber \\
          &   &     \times {\bar {\psi}}(p_4) \gamma_5 \psi(p_2).
\end{eqnarray}

{\bf{(vii) Nucleon intermediate state and the exchange of heavy axial meson}}

\begin{eqnarray}
 A_N^{A} (c) & = & g_{A} \left( \frac{f}{m_\pi} \right)
                 \frac{1}{q^2 - m_A^2} \frac{1}{p_i^2 - m_\pi^2} 
                        (2p_4 p_\pi + m_\pi^2) \nonumber \\
             &   & \times {\bar {\psi}}(p_4)\gamma^\mu b_\mu
              -2m_\pi {\bar {\psi}}(p_4)\gamma^\nu p_{\pi \nu} \gamma^\mu b_\mu
                 \psi(p_2),
\end{eqnarray}
where
\begin{eqnarray}
b_\mu & = & \bar {\psi}(p_3) \gamma_5 \gamma_\mu \psi(p_1)
\end{eqnarray}
The Numerical evaluation of the amplitudes written in the forms given above
can be carried out very effectively by using the techniques of the
Clifford algebra.

\section*{Appendix B}
In this appendix we present the discussion on the form of the
delta isobar propagator.

The free Lagrangian density for the massive spin-$3/2$ field
is written as \cite{43}
\begin{eqnarray}
{\cal L}_{\Delta} & = & \bar \Psi^{\mu} \Lambda_{\mu \nu} \Psi^{\nu},
\end{eqnarray}
where the most general form of $\Lambda_{\mu \nu}$ is given by \cite{43}
\begin{eqnarray}
\Lambda_{\mu \nu} & = & -[(-i\partial_\mu \gamma^\mu + m_\Delta)g_{\mu \nu}
           -iZ_1(\gamma_\mu \partial_\nu + \partial_\mu \gamma_\nu) \\ \nonumber
   &&      -Z_2\gamma_\mu(\partial_\lambda \gamma^\lambda) \gamma_\nu
           -Z_3 m_\Delta \gamma_\mu \gamma_\nu]
\end{eqnarray}
Here $Z_1$ is an arbitrary parameter subject to the restriction that 
$Z_1 \neq - 1/2, $\linebreak and $Z_2$ and $Z_3$ are defined as
\begin{eqnarray}
Z_2 = \frac{1}{2}(3Z_1^2+2Z_1+1),\; Z_3 = (3Z_1^2+3Z_1+1)
\end{eqnarray}
Physical properties of the free field do not
depend on the parameter $Z_1$,
which is chosen to be real. This is due to the fact that ${\cal L}_\Delta$
is invariant under the point transformation \cite{43}
\begin{eqnarray}
\Psi^\mu & \rightarrow & \Psi^\mu + a \gamma^\mu \gamma^\nu \Psi_\nu \\
A & \rightarrow & \frac{Z_1 - 2a}{1 + 4a},
\end{eqnarray}
where $a \neq -\frac{1}{4}$, but is otherwise arbitrary.

The wave equation for the spin-$3/2$ particle (the Rarita-Schwinger)
equation is written as
\begin{eqnarray}
\Lambda_{\mu \nu}(p) \Psi^\nu  =  0
\end{eqnarray}
Operating on Eq. (58) with $\gamma_\mu$ and $\partial^\mu$ we get the
local wave equation for a spin-$3/2$ particle along with the
constraint equations
\begin{eqnarray}
(i\partial_\nu - m_\Delta) \Psi^\mu & = & 0 \\
\gamma_\mu \Psi^\mu & = & 0 \\
\partial_\mu \Psi^\mu & = & 0
\end{eqnarray}
It should be noted that Eqs. (59) - (61) are obtained only when
the restriction $Z_1 \neq -1/2$ and the definitions of $Z_2$ and $Z_3$ as
given in Eq. (55) are used.

The propagator for a massive spin-$3/2$ particle satisfies the
following equation in the momentum space
\begin{eqnarray}
\Lambda_{\mu \nu} (p) G_{\Delta \alpha}^\nu & = & -g_{\mu \alpha}
\end{eqnarray}
Solving for G and making the particular choice of $ Z_1 = -1$ we get the
following equation for the delta isobar propagator
\begin{equation}
G_{\mu \nu}^\Delta (p)  =  -\frac{i(p\!\!\!/ + m_\Delta)}{p^2 - m_\Delta ^2}
                    [g_{\mu \nu} - \frac{1}{3}\gamma_\mu \gamma_\nu -
                    \frac{2}{3m_\Delta^2} p_\mu p_\nu + \frac{1}{3m_\Delta^2}
                   ( p_\mu \gamma_\nu - p_\nu \gamma_\mu )]
\end{equation}

A slightly different form of the $\Delta$ propagatator has been
suggested by Williams \cite{20}, which is given by
\begin{equation}
G_{\mu \nu}^{\Delta W} (p)  =  -\frac{i(p\!\!\!/ + m_\Delta)}{p^2 - m_\Delta ^2}
                    [g_{\mu \nu} - \frac{1}{3}\gamma_\mu \gamma_\nu -
                    \frac{2}{3p^2}p_\mu p_\nu + \frac{1}{3p^2}
              p_\eta \gamma^\eta ( p_\mu \gamma_\nu - p_\nu \gamma_\mu )]
\end{equation}
Williams has argued that the Eq. (64) should be the valid
form of the $\Delta$ isobar propagator both on and off the mass shell.
However, we show in the following that it can not be the correct spin-$3/2$
propagator.

First, we note that if $G_{\mu \nu}^{\Delta W}$ is the propagator
for a spin-$3/2$ particle, then it should satisfy
\begin{eqnarray}
\Lambda^{\prime \mu \nu} (p) G_\nu^{\alpha \Delta W} & = & -g^{\mu \alpha},
\end{eqnarray}
where $\Lambda^\prime$ is defined as
\begin{eqnarray}
\Lambda_{\mu \nu}^\prime & = & -[(-i\partial_\mu \gamma^\mu + m_\Delta)g_{\mu \nu}
           +i \lambda \gamma_\mu \partial_\nu -i \lambda \partial_\mu \gamma_\nu],
\end{eqnarray}
with the limit that $\lambda \rightarrow \infty$ (known as Feynman gauge).
However, Eq. (66) is not consistent with the general form of the
Lagrange function of a spin-$3/2$ particle as given by
Eqs. (54) and (55).

Second, in Ref. \cite{22}, it has been shown that the propagator
$G_{\mu \nu}^{\Delta W}$ has no inverse, and thus it can not be the
propagator of a spin-$3/2$ particle. A similar difficulty is associated
with the propagator suggested by Adelseck et al. \cite{21}.

In Fig. 9a and 9b, we study the effect of using the two delta isobar propagators
$G_{\mu \nu}^\Delta$ and $G_{\mu \nu}^{\Delta W}$ in the calculations of the
total and triple differential cross sections 
for the $p(p,n \pi^{+})p$ reaction to stress further the difficulties 
associated with the Williams propagator.
The solid lines show the results obtained by using the
former propagator while the dashed line the later one. It is clear that
the $G_{\mu \nu}^{\Delta W}$ produces very large cross sections
in a certain range
of beam energies due to the fact that the factor $p_i^2$ present in the
demoninators of some terms of this propagator, becomes
very small. This effect
is particularly very strong for the pre-emission graphs. For example, for
such a graph which is analogous to Fig. 1a we have
\begin{eqnarray}
p_i^2 & = & (p_2^2 - p_\pi^2) \nonumber \\
      & = &  m_N^2 - 2E_\pi m_N + m_\pi^2
\end{eqnarray}
It can be seen that for $E_\pi \simeq 0.479 \; GeV$ , $p_i$ is very close
to zero. This energy corresponds to an incident proton energy of
approximatly of 1.2 $GeV$. Therefore, at incident energies around this
value the terms which are proportional to
$(1/p_i^2)$ in $G_{\mu \nu}^{\Delta W}$ become very large in
the pre-emision graphs which in turn leads to huge total cross sections.
In the post-emission graphs this problem is not so severe.  This very clearly
shows the difficulty that one encounters while using the propagator
$G_{\mu \nu}^{\Delta W}$. It must be mentioned here that
normally the contributions
of pre-emission graphs are much smaller than those of the
post-emission ones if one uses the propagator $G_{\mu \nu}^\Delta$.

The effect of using $G_{\mu \nu}^{\Delta W}$ is very drastic in case of triple
differential cross sections as can be seen in Fig.
9b. Williams propagator leads to
a very large cross section (which looks almost like a pole) at certain value
of the outgoing pion momenta due to the same reason as discussed above. 

\end{appendix}

\newpage
\clearpage
\begin{table}
\caption {\bf {Coupling constants for the $NN \, meson$
vertices used in the calculations}}
\vspace{1.1cm}
\begin{tabular}{|c|c|c|c|c|} \hline
 Meson & $g^2/4\pi$ & $\ell$ & $\Lambda$ & mass \\
       &             &        & (\footnotesize{GeV} ) & (\footnotesize{GeV})
 \\ \hline
$\pi    $ & 12.562 & 0.1133 & 1.005 & 0.138 \\
$\sigma $ & 2.340  & 0.1070 & 1.952 & 0.550 \\
$\omega $ & 46.035 & 0.0985 & 0.984 & 0.783 \\
$\rho   $ & 0.317  & 0.1800 & 1.607 & 0.770 \\
$k_{\rho}$ = 6.033, $k_{\omega}$ = 0.0, $g_{\rho \pi \pi} = 2g_\rho$ & & & & \\ \hline
\end{tabular}
\end{table}
\newpage
\clearpage
\begin{table}
\caption{\bf{Isospinfactors for pole diagramms}}
\vspace{1.1cm}
\begin{minipage}[t]{0.40\textwidth}
 \centering
\begin{tabular}{|c|c|c|} \hline
\multicolumn{3}{|c|}{\it \bf nucleon pole}\\   \hline
\multicolumn{3}{|c|}{$pp \rightarrow np\pi^+$}\\ \hline
graph & isovector & isoscalar \\ \hline
a    & $\sqrt{2}$ & $\sqrt{2}$ \\
b &0&0\\
c&0&0\\
d&$\sqrt{2}$&$\sqrt{2}$\\
pre a &$-\sqrt{2}$ &$\sqrt{2}$\\
pre b &$2\sqrt{2}$ &0\\
pre c &$2\sqrt{2}$ &0\\
pre d &$-\sqrt{2}$ &$\sqrt{2}$\\ \hline
\multicolumn{3}{|c|}{$pp \rightarrow pp\pi^o$}\\ \hline
\multicolumn{2}{|c|}{all graphs} & 1\\ \hline
\end{tabular}
\end{minipage}
\begin{minipage}[t]{0.40\textwidth}
\centering
\begin{tabular}{|c|c|c|} \hline
\multicolumn{3}{|c|}{\it \bf delta pole}\\   \hline
\multicolumn{3}{|c|}{$pp \rightarrow np\pi^+$}\\ \hline
graph & isovector & isoscalar \\ \hline
a    & $-\sqrt{2}/3$ & 0 \\
b &$\sqrt{2}$&0\\
c&$\sqrt{2}$&0\\
d&$-\sqrt{2}/3$& 0\\
pre a &$\sqrt{2}/3$ &0\\
pre b &$\sqrt{2}/3$ &0\\
pre c &$\sqrt{2}/3$ &0\\
pre d &$\sqrt{2}/3$ &0\\ \hline
\multicolumn{3}{|c|}{$pp \rightarrow pp\pi^o$}\\ \hline
\multicolumn{2}{|c|}{all graphs} & 2/3\\ \hline
\end{tabular}
\end{minipage}
\end{table}
\begin{table}
\caption{\bf{Isospinfactor for direct diagramms}}
\vspace{1.1cm}
\centering
\begin{tabular}{|c|c|c|} \hline
\multicolumn{3}{|c|}{\it \bf intermediate}\\  \hline
\multicolumn{3}{|c|}{$pp \rightarrow np\pi^+$}\\ \hline
graph & \multicolumn{2}{|c|}{}  \\ \hline
e    & \multicolumn{2}{|c|}{-i$\sqrt{2}$ } \\
f &\multicolumn{2}{|c|}{i$\sqrt{2}$}\\
g&\multicolumn{2}{|c|}{i$\sqrt{2}$}\\
h&\multicolumn{2}{|c|}{-i$\sqrt{2}$}\\ \hline
\multicolumn{3}{|c|}{$pp \rightarrow pp\pi^o$}\\ \hline
\multicolumn{2}{|c|}{all graphs} & 0\\ \hline
\end{tabular}
\end{table}
\clearpage
\newpage
\begin{center}{Figure Captions} \end{center}
\begin{itemize}
\item [Fig. 1].
Feynman diagrams for emission of a pion in the nucleon -
nucleon collisions. (a) Exchanged meson starts from nucleon 2 (momentum
$p_2$), is absorbed by nucleon 1 (momentum $p_1$) which is excited to the
nucleon or delta isobar intermediate state with momentum ($p_i$)
which then decays into
the nucleon 3 (momentum $p_3$) and the outgoing pion (momentum $p_\pi$). The
nucleon 2 goes on to nucleon 4 with momentum ($p_4$). (b) Exchange part
of the diagram (a). (c) Same as (a) but the exchanged meson starts from
nucleon 1 and is captured by nucleon 2 which is excited to
intermediate states which then decay into nucleon 4 and the outgoing
pion. Nucleon 1 goes on to become nucleon 3. (d) Exchange part of diagramme
(c). (e)-(h) Direct and exchange diagrams showing the processes
where the exchanged $\rho$ meson starting from one of the interacting nucleons
decays into two pions in flight. One of them is the outgoing
pion, the another one is absorbed by the other nucleon. These diagrams are
referred as intermediate graphs in the text. Note that there are also the
pre-emission counter-parts of the diagramms (a) - (d) where pion are
emitted before collisions. These graphs are not shown here, but their
contributions are included in the calculations.

\item[Fig.2a].
The total cross section for the $p(p, n\pi^{+}) p$ 
reactions as a function of beam
energy.  The dashed and dashed-dotted lines
represent the results of the
calculations performed with only delta isobar intermediate states and
only nucleon intermediate states, respectively. The sum of all the
graphs is represented by the solid curve. The experimental data are 
taken from \cite{3}.

\item[Fig2b].
Same as Fig. 2a but for the reaction $p (p, p\pi^{0}) p$.

\item[Fig. 3].
Contributions of various exchanged mesons to the total
cross section for the reaction $p(p, p \pi^{+})p$
as a function of beam energy. The long dashed, dotted, short dashed
and dashed-dotted curves represent the contributions of
$\pi$ exchange, $\rho$ exchange, $\omega$ exchange
and $\sigma$ exchange, respectively.

\item[Fig. 4].
Total cross sections for the reaction
$p(p, p \pi^{+})p$
as a function of beam energy for
two values of the cut-off parameter of the $\pi NN$ vertex.
The solid (long dashed) line is the result of calculations
performed with a value of 1.005 GeV (0.631 GeV)
for this parameter.

\item[Fig. 5].
Ratio of the total cross sections calculated with
pseudovector and pseudoscalar couplings for the $\pi NN$ vertex for the
same reaction as in Fig. 4, as a function of beam energy. The long dashed line
represents the results obtained with only nucleon intermediate states
while the solid line contain all the graphs.

\item[Fig. 6].
The double differential cross section for the reaction
$pp \rightarrow \pi^{+} + X $ as a function of pion momentum for pion
angles of $20^0$, $40^0$ and $60^0$ at a beam energy of 800 $MeV$.
The dashed (dotted) line represents the
results obtained with only delta isobar (only nucleon) intermediate states. 
The solid line shows the
results obtained by including all the graphs.

\item[Fig. 7].
The triple differential cross sections for the
$p(p, n\pi^{+})$ reaction at the beam energy of 800 $MeV$ as a
function of the outgoing pion momentum. The upper part shows the
results for the proton and pion angles of $15^0$ and $21^0$ respectively
while the lower part for $25^0$ and $40^{0}$ respectively. The dashed lines
show the results when only delta isobar intermediate states are included
in the calculations while the solid lines depict the
results obtained by including all the graphs.

\item[Fig. 8].
The effect of using various forms of delta decay width
in the calculation of triple differential cross sections
for the same reaction as in Fig. 7 and at the same beam energy.
The solid and dashed lines
represent the results obtained with the forms of the 
delta decay widths given
by Verwest and Dmitriev with a free delta decay widths of 120 MeV.
The dotted lines
represent the results obtained with the later but with a 
a free delta decay width of 100 MeV.
The results are shown as a function of outgoing pion momentum.

\item[Fig. 9a].
Total cross section
for the $p(p, p\pi^{+})p$ reaction
as a function of beam energy
calculated with the propagators of
Benmerrouche et al. \cite{22} (solid line) and Williams \cite{20}
(long dashed line).

\item[Fig. 9b].
Triple differential cross section for the $p(p, p\pi^{+}) p$ reaction
as a function of outgoing proton momentum corresponding to proton and
pion angles of 15$^\circ$ and 21$^\circ$ respectively, The solid and 
long dashed curves have the same meaning as in Fig. 9a.

\end{itemize}

\end{document}